\documentclass[11pt]{article}
\usepackage{amssymb}
\usepackage{epsfig}
\usepackage{graphicx,graphics,color}

\usepackage{multirow}
\def\Br{\mathrm{BR}}

\def\be{\begin{equation}}
\def\ee{\end{equation}}
\def\bea{\begin{eqnarray}}
\def\eea{\end{eqnarray}}

\begin{document}
\setcounter{footnote}{0}
\vspace*{-2cm}
\begin{flushright}
LPT Orsay 13-76
\vspace*{2mm}							
\end{flushright}

\begin{center}
\vspace*{15mm}

\vspace{1cm}
{\Large\bf 
Neutrino Physics, Lepton Flavour Violation and the  LHC}\\
\vspace{1cm}

{\bf A. Abada }

\vspace*{.5cm} 
 Laboratoire de Physique Th\'eorique, CNRS -- UMR 8627, \\
Universit\'e de Paris-Sud 11, F-91405 Orsay Cedex, France
\vspace*{.2cm} 

\end{center}
\vspace*{10mm}
\begin{abstract}
We will briefly review flavour violation in the lepton sector: starting from neutrino oscillations and their implications, 
we consider several charged lepton flavour violating observables at high and low energies.
 We present new physics models 
 and discuss the r\^ole of the latter in disentangling them.
In particular, we show how the interplay of different observables allows to 
derive important information on the underlying mechanism of lepton
flavour violation.
As an example,  we discuss the impact of a type-I SUSY seesaw concerning lepton
flavour violation  at low energies and at colliders (LHC and a future Linear Collider). 
\end{abstract}

\vspace*{3mm}
%

\section{Introduction}
Neutrino ($\nu$) oscillation experiments provide indisputable evidence for flavour
violation in the neutral lepton sector. This is manifest in leptonic charged currents 
and, as it occurs for the quark sector, the corresponding amount of  flavour violation can be encoded into a mixing matrix, 
the $U^{\mathrm{PMNS} }$. Nevertheless, oscillation data only adds to the Standard Model's fermion flavour puzzle: the 
mixing patterns on the quark and lepton sectors are very different; available data on neutrino masses
(mass squared differences and bounds on the absolute mass scale) suggests that if these are to be accommodated 
via a minimal extension of the Standard Model (SM) by additional right-handed (RH) neutrinos, the Yukawa couplings of the corresponding 
Dirac mass term ($\sim Y^\nu \nu_L \nu_R $) would be $\mathcal{O}(10^{-12})$, rendering even worse the 
fermion hierarchy puzzle.

In order to accommodate $\nu$ oscillation data, the SM must clearly be extended, and additional degrees of freedom incorporated (RH neutrinos, triplet Higgs, ...). Moreover, should neutrinos be Majorana fermions, 
new physics (NP) at scales much lighter or far heavier than the electroweak (EW) one can be present. 
Current experimental (and cosmological) data on neutrino phenomena leaves numerous questions unanswered, such 
as the hierarchy in the neutrino spectrum, leptonic CP violating phases and their possible r\^ole in 
the evolution of the early Universe, unitarity violation in neutrino interactions, non-standard interactions, among many 
others. Moreover current reactor and accelerator anomalies suggest that there might be extra fermionic gauge singlets 
(sterile states). This would in turn imply that instead of a 3-neutrino mixing scheme we could have 3+1 (or more)-mixing schemes,  and this would have considerable implications on NP models. 
Possible SM extensions aiming at incorporating massive neutrinos and leptonic mixing will also open the door 
to many new phenomena, such as flavour violation in the charged lepton sector~(cLFV), leptonic electric dipole 
moments, contributions to the muon anomalous magnetic moment, .... The new states present in these  
extensions can also give rise to interesting collider signatures.  

Other than $\nu$-mixings, there are several observational problems and theoretical caveats  suggesting that 
NP is indeed required: the former are related to the baryon asymmetry of the Universe and the need for a 
dark matter candidate, while among the latter one can mention the hierarchy problem, the flavour puzzle, or fine-tuning 
in relation to EW symmetry breaking. There are numerous well motivated and appealing models of NP that aim at addressing these issues, and which are currently being actively investigated and searched for. 
Disentangling the NP model and in particular, probing the underlying mechanism of neutrino mass generation, 
requires investigating all available observables, arising from all fronts - high-intensity, high-energy and 
cosmology - as well as thoroughly exploring the interplay between them. 

Here, we will focus on charged lepton flavour violating observables at low-energies and at colliders, and we discuss 
their powerful r\^ole in shedding light on the NP model. We will begin by briefly reviewing the effective approach 
(model-independent) and subsequently consider different SM extensions. Finally, and as an example, we will address in some detail 
cLFV in the supersymmetric (SUSY) seesaw.

\section{cLFV observables}
\begin{figure}[h!]
\begin{center}
\begin{tabular}{ccc}
\raisebox{3mm}{
\includegraphics[height=25mm, width=35mm]{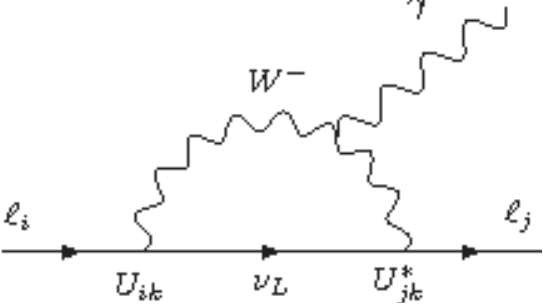}
}
\hspace*{5mm} & \hspace*{5mm}
\raisebox{1mm}{\includegraphics[height=25mm, width=35mm]{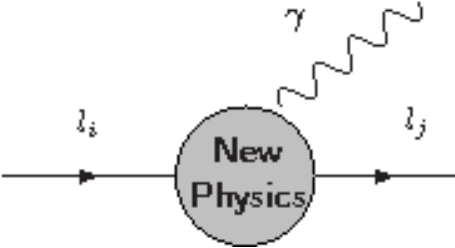}}
\hspace*{5mm} &\hspace*{5mm}
\raisebox{2.7mm}{\includegraphics[height=26mm, width=45mm]{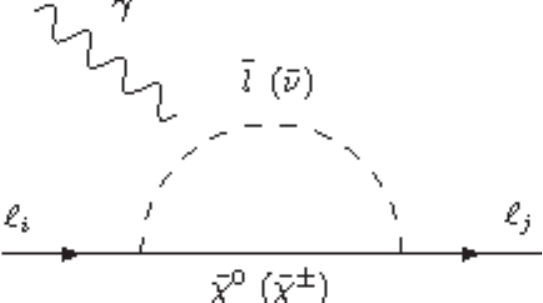}} \vspace*{-1mm}\\
\vspace*{-1mm}(a) \hspace*{5mm}&\hspace*{5mm} (b) \hspace*{5mm}&\hspace*{5mm} (c)
\end{tabular}
\caption{ 
\small Radiative decays $\ell_i \to \ell_j \gamma$: (a) in the SM ($m_\nu\neq 0$), in NP models (b) 
and in SUSY models (c). }\label{LFVBSM}
\end{center}
\end{figure}

In the SM, as it was originally formulated (massless $\nu$), 
cLFV processes are strictly forbidden. 
With massive $\nu$'s (no assumption being made on the 
mechanism of $\nu$ mass generation), 
cLFV processes are suppressed by 
the tiny $\nu$ masses (GIM-like suppression):
for example, the branching ratio (BR) of the $\mu \to e \gamma$ decay, schematically depicted in Fig.~\ref{LFVBSM}(a), is given by~\cite{Petcov:1976ff}
\begin{equation}\label{eq:SM:BRmue}
\mathrm{BR}(\mu \to e \gamma) = \frac{3 \alpha}{32 \pi} \left|   
\sum_i U^*_{\mu i} U_{e i} \frac{m^2_{\nu_i}}{M_W^2}
\right|^2\,,
\end{equation}
and using known oscillation parameters 
($U=U^{\mathrm{PMNS} }$ the leptonic mixing matrix)~\cite{Beringer:1900zz},
one finds 
BR$(\mu \to e \gamma)\lesssim 10^{-54}$, 
thus unaccessible to present and future experiments! In contrast with flavour violation  in the hadronic sector, where
phenomena such as neutral meson mixings and rare decays can be
successfully explained by the SM, 
cLFV  signals indisputably the presence of new physics.  Indeed, 
the additional  particle content and the new flavour dynamics present 
in many extensions of the SM may give contributions to 
cLFV processes such as radiative 
(e.g. $\mu \to e\gamma$) and three-body decays  
(e.g. $\mu \to eee$), so that 
the observation of such processes would provide an unambiguous 
signal of NP, see Fig.~\ref{LFVBSM}(b).

So far, no signal of cLFV has been observed, 
even though the searches for rare leptonic decays have been an
important part of the experimental program for several decades.
 Currently, 
the search for manifestations of cLFV constitutes the goal of 
several
experiments 
 exclusively dedicated to look for signals of  rare lepton decay 
processes. Equally interesting LFV observables are $\mu-e$ conversion in heavy
nuclei: although significant improvements are expected regarding the
experimental sensitivity to $\mu \to e \gamma$ ($ <
10^{-13}$)~\cite{Adam:2013mnn}, the most challenging experimental
prospects arise for the CR($\mu-e$) in heavy nuclei such as titanium
or gold.
The possibility of improving the sensitivities to values as
low as $ \sim 10^{-18}$ renders this observable an extremely powerful
probe of LFV in the muon-electron sector. 
In Table~\ref{table:LFV:bounds}, 
we  briefly survey some of the current bounds for the different cLFV
processes, as well as the future experimental sensitivity. 
{
\begin{table}[h!]
\begin{center}
\begin{tabular}{|l|c r|c r|}
\hline
LFV process & Present bound & & Future sensitivity & \\
\hline
BR($\mu \to e \gamma$) & $5.7\times 10^{-13}$&
\cite{Adam:2013mnn}&$6\times 10^{-14}$& \cite{Baldini:2013ke} \\ 
BR($\tau \to e \gamma$) &$3.3\times 10^{-8}$ & \cite{Aubert:2009ag}&
$ 10^{-9}$& \cite{Meadows:2011bk} \\ 
BR($\tau \to \mu \gamma$) & $4.4\times 10^{-8}$&
\cite{Aubert:2009ag}&$ 10^{-9}$ & \cite{Meadows:2011bk} \\ 
\hline
BR($\mu \to 3 e $) &$1.0 \times 10^{-12}$ &
\cite{Bellgardt:1987du}&$\sim 10^{-16}$  
&  \cite{Blondel:2013ia}\\ 
BR($\tau \to 3 e $) & $2.7 \times 10^{-8}$& \cite{Hayasaka:2010np}&$2.3
\times 10^{-10}$ & \cite{Meadows:2011bk} \\ 
BR($\tau \to 3 \mu$) & $2.1 \times 10^{-8}$& \cite{Hayasaka:2010np}&$8.2
\times 10^{-10}$ & \cite{Meadows:2011bk} \\ \hline
CR($\mu-e$, Ti) & $4.3 \times 10^{-12}$& \cite{Dohmen:1993mp}&
${\mathcal{O}}(10^{-18})$ &
\cite{Barlow:2011zza} \\ 
CR($\mu-e$, Au) & $7 \times 10^{-13}$& \cite{Bertl:2006up}& 
                                 
& 
 \\
\hline 
CR($\mu-e$, Al) && & ${\mathcal{O}}(10^{-16})$&  \cite{Cui:2009zz}\\\hline
\end{tabular}
\end{center}
\caption{\small Present bounds and future sensitivities for several LFV
  observables.}
\label{table:LFV:bounds}
\end{table}
}

In addition to low-energy experiments, there are also searches for cLFV
at high-energies: the presence of new flavour violating physics 
can be directly signaled via 
the LFV production and/or  decays of heavy states 
(which must be  nevertheless sufficiently light to be produced at the LHC or a future Linear Collider). Moreover, 
data from LHCb is also expected to directly constrain LFV (as well as lepton number violation) 
in meson and in tau-lepton decays.

In the absence of cLFV  and other signals, one  could constraint the parameter space of NP models (scale, couplings, ...) and this may directly probe the neutrino mass generation mechanism. On the other hand, if cLFV is indeed observed, then one should compare the signal with peculiar features of a given model (predictions of observables, patterns of correlations between observables). 
 A possibility to address these NP scenarios is to use the effective approach and study a given (cLFV) observable  in a model independent way.

\section{cLFV: the effective approach}
In the SM, lepton number is an accidental symmetry due to the gauge
group and particle content. The generation of $\nu$ masses 
exclusively using the SM field content requires adding  non-renormalisable operators of dimension 5 (or
higher), that break lepton number, to the 
SM Lagrangian. 
Independent of the model, the only possible $d=5$
operator is the Weinberg operator,  
$
\delta{\mathcal L}^{d=5} = \frac{1}{2}\, c_{\alpha \beta}^{d=5} \,
\left( \overline{\ell_L^c}_{\alpha} \tilde \phi^* \right) \left(
\tilde \phi^\dagger \, {\ell_L}_{ \beta} \right) + {\rm h.c}.\, ,
$
where $\ell_L$ stands for the lepton doublets and 
and $\tilde \phi$ is related to the SM Higgs doublet. 
The coefficient $c^{d=5}$ is a matrix of 
inverse mass dimension, which is not invariant under  $B-L$,
 and is thus a source of Majorana $\nu$ masses. 
 
 Among the dimension 6 operators (second order in
an $1/ M$ expansion, $M$ being the scale of NP), one finds four-fermion operators responsible for cLFV processes.   
The breaking of lepton number, as required by Majorana $\nu$ masses, then 
provides a natural link between neutrino mass generation and cLFV. The effective Lagrangian then reads
\begin{equation}
\mathcal{L}_{\mathrm{eff}}= \mathcal{L}_{\mathrm{SM}} +
{\frac{1}{M}}{ c^{d=5} {\mathcal{O}}^{d=5} }+  
{\frac{1}{M^2}}{ c^{d=6} {\mathcal{O}}^{d=6} } + \cdots\,, 
\end{equation}
where the NP, valid at the scale $M$ is encoded in the coefficients, $c^{d>4}$. 

\begin{figure}[h!]
\begin{center}
\begin{tabular}{ccc}
\raisebox{3mm}{\includegraphics[height=23mm, width=34mm]{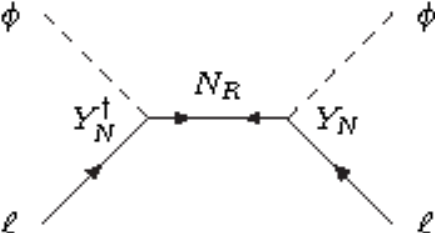}}
\hspace*{5mm} & \hspace*{5mm}
\raisebox{3mm}{\includegraphics[height=27mm, width=24mm]{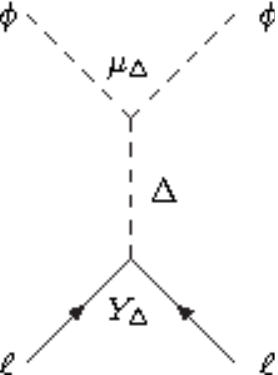}}
\hspace*{5mm} &\hspace*{5mm}
\raisebox{0mm}{\includegraphics[height=24mm, width=34mm]{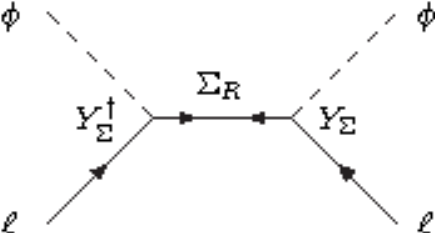}} \vspace*{-3mm}\\
 \vspace*{-3mm}
(a) \hspace*{5mm}&\hspace*{5mm} (b) \hspace*{5mm}&\hspace*{5mm} (c)
\end{tabular}
\caption{
\small Seesaw mechanisms: (a) singlet fermion,  (b) triplet fermion
and (c) triplet scalar exchange.} 
\label{fig:seesawI-III}
\end{center}
\end{figure}

In the case of a minimal extension of the SM by heavy fields, it  can be shown that one can only have  three
types of basic seesaw mechanisms, depending on the nature of the new
heavy fields: right-handed neutrinos (type I), 
 heavy scalars (type II) 
 or 
fermionic triplets (type III), 
 as depicted in
Fig.~\ref{fig:seesawI-III}. It is important to notice that all 
these mechanisms can be embedded into larger frameworks such as grand unified theories (GUTs), SUSY and  extra dimensions. 

 In the case of a fermionic (type I or III) seesaw, 
the heavy propagator is expanded as 
$\frac{1}{ D\!\!\!\!/ \ -M}\,\sim\, -{\frac{1}{M}}\,-\, \frac{1}{M}\, D\!\!\!\!/
\frac{1}{M}+\cdots$
The first term induces a $d=5$ scalar
operator, which flips chirality and generates a $\nu$-mass term. 
The second term ($\sim 1/M^2$) preserves chirality and
induces a correction to the kinetic terms of the light fields. 
The coefficients 
$ c^{d=6}\propto   {1\over M^2}$ are
suppressed compared to those associated to the $d=5$ operator, 
$c^{d=5}\propto  {  {1\over M}}$.  
The situation is different in the case of heavy scalar triplets,
since the scalar propagator expands as 
$\frac{1}{D^2 -M^2}\,\sim\, -{\frac{1}{M^2}}\,-\, \frac{D^2}{M^4} +\ \cdots$, 
implying that the $d=5$ operator already scales as $1/M^2$. 
In the case of the type I and III seesaws,
$c^{d=5}=Y_{N}^{T}\frac{1}{M_{N}}Y_{N}$, $Y_{N}$ being
the Yukawa couplings to the Higgs field 
and $M_N$ the heavy fermion masses. 
Accommodating $\nu$ data with natural coefficients ($c^{d=5} \sim {\mathcal{O}}(1)$), implies $ M_N\sim 10^{15}$ GeV,
intriguingly close to GUT scale. However,  
the scale of NP can be lowered to $\sim 1 $ TeV 
if one allows for couplings as small as the charged lepton ones.
In the case of a scalar triplet, $c^{d=5} \propto Y_\Delta
\mu_\Delta/M^2_\Delta$: the scale $\mu_\Delta$ can be directly related
to the smallness of $m_\nu$ thus allowing to have $M_\Delta \sim $ TeV
with natural Yukawa couplings.

Since such a $d=5$ operator is
characteristic to all models with Majorana $\nu$, 
the coefficient $c^{d=5}$ does not allow to
discriminate among the different models. In order to do so, 
one must 
either produce the heavy mediators or call upon the low-energy effects
of the different  $d=6$ operators. There is a large number   
of such operators 
but here we will only 
focus on those inducing cLFV processes. On Table~\ref{table:operators} we list the  $d=6$ 
cLFV operators as well as the corresponding coefficient 
for each type of seesaw and, for comparison, the $d=5$ coefficient.  
\begin{table}[!h]\renewcommand{\arraystretch}{1.}
{
\begin{center}
\begin{tabular}{|c||c|c|c|}
\hline
& \multicolumn{3}{c|}{Effective Lagrangian
  $\mathcal{L}_{eff}=c_{i}\mathcal{O}_{i} $} \rule[-5 pt]{0pt}{18
  pt}\\ 
\cline{2-4}
Model & $c^{d=5}$ & $c^{d=6}$ & $\mathcal{O}^{d=6}$  \rule[-5
pt]{0pt}{18 pt}\\ 
\hline 
\hline
$\begin{array}{c}
\mathrm{Fermionic \,Singlet} \vspace*{-1mm}\\
\mathrm{(type \,I)}
\end{array}$
& $Y_{N}^{T}\frac{1}{M_{N}}Y_{N}$ &
$\left(Y_{N}^{\dagger}\frac{1}{M_{N}^\dagger}\frac{1}{
    M_N}Y_{N}\right)_{\alpha\beta}$ &
$\left(\overline{\ell_{L\alpha}}\widetilde{\phi}\right)i\partial\!\!\!/
\left(\widetilde{\phi}^{\dagger}\ell_{L\beta}\right)$ 
\rule[-14 pt]{0pt}{34 pt}\\ 
\hline
$\begin{array}{c}
\mathrm{Scalar \,Triplet } \vspace*{-1mm}\\
\mathrm{(type \,II)}
\end{array}$
 & $4Y_{\Delta}\frac{\mu_{\Delta}}{M_{\Delta}^{2}}$ &
$\frac{1}{ M_\Delta^{2}}Y_{\Delta \alpha\beta}Y_{\Delta
  \gamma\delta}^{\dagger}$ &$
\left(\overline{\widetilde{\ell_{L\alpha}}}
\overrightarrow{\tau}\ell_{L\beta}\right)\left(\overline{\ell_{L\gamma}}
\overrightarrow{\tau}
  \widetilde{\ell_{L\delta}}\right)$
\rule[-14 pt]{0pt}{34 pt}\\ 
\hline
$\begin{array}{c}
\mathrm{Fermionic \,Triplet} \vspace*{-1mm}\\
\mathrm{(type \,III)}
\end{array}$
 & $Y_{\Sigma}^{T}\frac{1}{M_{\Sigma}}Y_{\Sigma}$ &
$\left(Y_{\Sigma}^{\dagger}\frac{1}{M_{\Sigma}^\dagger}\frac{1}{
    M_\Sigma}Y_{\Sigma}\right)_{\alpha\beta}$ &
$\left(\overline{\ell_{L\alpha}}\overrightarrow{\tau}
\widetilde{\phi}\right)iD\!\!\!\!/\left(
\widetilde{\phi}^{\dagger}\overrightarrow{\tau}\ell_{L\beta}\right)$ 
\rule[-14 pt]{0pt}{34 pt}\\ 
\hline
\end{tabular}
\end{center}
\caption{\small \mbox{Dimension 6 operators (and coefficients) responsible for cLFV and  corresponding $d=5$ coefficients.}}
\label{table:operators}
}\renewcommand{\arraystretch}{1.0}
\end{table}\\
From a symmetry point of view, it is natural to have large $c^{d=6}$ 
coefficients, since the $d=6$ operators preserve $B-L$, 
in contrast with the $d=5$ operator. For example, 
in the type II seesaw, the dimensionfull $\mu_\Delta$ coefficient, which is directly related to the smallness of $m_\nu$,
does not affect the dimension 6 operator. 
However, decoupling the $d=5$ and $d=6$ coefficients is not
possible in the fermionic seesaw (see   
Table~\ref{table:operators}). 

In the effective approach (for a review, see~\cite{Abada:2007ux}), the observables can be written in
terms of effective parameters, encoding the flavour mixing
generated by the model, which remains valid  up to a scale $\Lambda$. For example, in models onto which a type II seesaw is embedded ($\Lambda=M_\Delta$),
the BRs for radiative and three body decays read 
\begin{equation}\label{eq:BR:eff}
\frac{\textrm{BR}(l_{i}\rightarrow l_{j}\gamma)}
{\textrm{BR}(l_i\rightarrow l_{j} \nu_i \bar{\nu}_j)}
={\alpha \over 48 \pi} 
{25\over 16} \frac{\left|{\Sigma_k}{Y_\Delta}_{l_kl_{i}}^{\dagger}
{Y_\Delta}_{l_{j}l_k}\right|^{2}}{G_{F}^{2}\Lambda^4}\, \,;\quad
\textrm{BR}(\ell_l^-\rightarrow l_i^+ l_j^- l_j^-)
= {1\over  { G_F^2\Lambda}^{4}} |Y_{\Delta_{l i} }|^2 |Y_{\Delta_{jj}}|^2\,.
\end{equation}
\noindent A signal from MEG (Table \ref{table:LFV:bounds})
i.e.  $10^{-14} \lesssim{\rm BR}(\mu \to e \gamma) 
\lesssim 10^{-13}$ will put constraints on 
$\Lambda$, depending on the size of the couplings:  assuming 
natural values of $Y_{\Delta}\sim \mathcal{O}(1)$ implies 
that 15~TeV $< \Lambda < $ 50~TeV, while for $Y_{\Delta}\sim
\mathcal{O}(10^{-2})$, BR$(\mu \to e \gamma)$ within MEG reach would 
lead to  0.15 TeV $< \Lambda < $ 0.5 TeV. If the triplet mass is of the order of the TeV, then one could expect some signals at the LHC, like the production of doubly charged Higgs triplet decaying into same sign leptons (a striking signal,  clean from any SM background). Collider signatures depend on the different parameters of the model and so far, negative LHC searches have already allowed to constrain the parameter space of this model~\cite{Melfo:2011nx}; moreover the existing synergy between high- and low-energy observables also allows to infer information on the neutrino mass spectrum and CP violating phases~\cite{Garayoa:2007fw}.

\section{cLFV and new physics}
Depending on the $\nu$ mass generation mechanism, 
 one can have very different scenarios of cLFV 
at low-energies. 
Whichever NP is called upon to explain the
origin of (Majorana) $\nu$ masses and mixings, whether or not it is sufficiently 
large to generate observable cLFV strongly depends on two main 
ingredients: the scale of
 NP (not necessarily the scale of the seesaw mediator)
and the amount of mixing present in the lepton sector parametrized by an effective mixing $\theta_{i j }$ ($U^{\mathrm{PMNS}}$ and additional mixings).

There are several classes of well-motivated extensions of the SM,
aiming at overcoming both its theoretical and experimental
shortcomings. These models can either offer new explanations for the
smallness of $m_\nu$  (e.g. through a geometrical suppression
mechanism, as is the case of large extra dimensions, or then R-parity
violation in the case of SUSY models), or onto them one can embed a
seesaw mechanism. In addition, these extensions can provide new
sources of LFV~\cite{Raidal:2008jk}. In general, the low-energy cLFV observables
can be significantly enhanced when compared to the minimal seesaw\footnote{For a discussion on observable cLFV 
	signals in TeV-scale type I seesaw and Higgs triplet models, see e.g.~\cite{Dinh:2013vya} and~\cite{Alonso:2012ji}
	and references therein.}. 

For instance, in the framework of enlarged Higgs sectors (as is the case of Little Higgs models~\cite{Blanke:2009am}), 
new couplings between SM leptons, new "mirror" fermions and heavy gauge bosons are sources of cLFV, being also at the origin of $m_\nu\ne 0$; this leads to scenarios of strongly correlated cLFV observables. However, such models require a sizable fine-tuning of the parameters as they typically induce excessively large contributions to low-energy cLFV observables. 

Another class of models generating non-trivial lepton flavour structures
is associated to the displacement of SM fermions 
along extra dimensions (scenarios with either large flat~\cite{ArkaniHamed:1999dc} or 
small warped~\cite{Gherghetta:2000qt} extra dimensions).  
For example, in the case of RS warped extra dimensions, compatibility with 
BR$(\mu\to~e \gamma)$ imposes that the mass of the Kaluza-Klein excitations
 be $M_{KK} \gtrsim 30$ TeV, thus beyond present
LHC reach.  Possible ways out include non-geometrical flavour structures, or an increase of the gauge 
symmetry~\cite{Iyer:2012db}.

Non-constrained SUSY extensions of the SM introduce new sources of LFV through (generic) soft-breaking SUSY terms, giving rise to flavour violating neutral and charged lepton-slepton vertices. Even if independent of the mechanism of neutrino mass generation, these sources induce sizable contributions to different cLFV observables (for a recent comparative discussion see~\cite{Arana-Catania:2013nha}). 

From these examples it is clear that NP models can predict (or accommodate) extensive ranges for the cLFV observables. 
Thus, in order to disentangle the underlying model of LFV, one has to explore the peculiar features (correlations of observables, etc.) of each model. This can be illustrated, as done in~\cite{Buras:2010cp}, by the following comparison of ratios of different low-energy cLFV observables for models such as Little Higgs (with T-parity), MSSM (Higgs or dipole dominance) and 4$^\mathrm{th}$ generation, as can be seen  in Table~\ref{ratios}.

 \begin{table}[ht]
  \renewcommand{\arraystretch}{1.1}
 \begin{center}
\begin{tabular}{|l|c|c|c|c|}
\hline
ratio & LHT  & \hspace*{1mm}MSSM (dipole) \hspace*{1mm}&\hspace*{1mm}
MSSM (Higgs)\hspace*{1mm}&SM4 \\\hline\hline 
{$\frac{\Br(\mu\to eee)}{\Br(\mu\to e\gamma)}$}  
& \hspace{.0cm}
{0.02\dots1}\hspace{0.cm}  &
{$\sim6\times 10^{-3}$}
&{$\sim6\times 
10^{-3}$} & $0.06\dots {2.2}$ \\ 
$\frac{\Br(\tau\to eee)}{\Br(\tau\to e\gamma)}$   &
0.04\dots0.4     &$\sim1\times 10^{-2}$ & ${\sim1\times 10^{-2}}$&
$0.07\dots  {2.2}$\\ 
$\frac{\Br(\tau\to \mu\mu\mu)}{\Br(\tau\to \mu\gamma)}$
&0.04\dots0.4     &$\sim2\times 10^{-3}$ & $0.06\dots0.1$& $0.06\dots
 {2.2}$ \\\hline 
$\frac{\Br(\tau\to e\mu\mu)}{\Br(\tau\to e\gamma)}$  &
0.04\dots0.3     &$\sim2\times 10^{-3}$ & $0.02\dots0.04$& $0.03\dots
 {1.3}$ \\ 
{$\frac{\Br(\tau\to \mu ee)}{\ \ \Br(\tau\to \mu\gamma)}$}  &
{0.04\dots0.3}
&{$\sim1\times 10^{-2}$} 
&{${\sim1\times 10^{-2}}$}&
$0.04\dots  {1.4}$\\ 
$\frac{\Br(\tau\to eee)}{\Br(\tau\to e\mu\mu)}$     &
0.8\dots2   &{$ {\sim5}$} & 0.3\dots0.5& 
$1.5\dots 2.3$\\ 
{$\frac{\Br(\tau\to
    \mu\mu \mu)}{\ \Br(\tau\to \mu e e)}$}
& {0.7\dots1.6}    
&{$\sim0.2$} &
{5\dots10}& $1.4 \dots 1.7$ \\\hline  
$\frac{ \mathrm{CR}(\mu \mathrm{Ti}\to e \mathrm{Ti})}{\Br(\mu\to e\gamma)}$
& \hspace*{1mm}$10^{-3}\dots 10^2$   \hspace*{1mm}  & $\sim 5\times
10^{-3}$ & $0.08\dots0.15$& \hspace*{1mm} $10^{-12}\dots
26\hspace*{1mm}$\\\hline 
\end{tabular}
\end{center}
\renewcommand{\arraystretch}{1.0}
 \caption{
 Comparison of various ratios of observables  in the LHT model,
 the MSSM without
and with significant Higgs contributions,
and the SM with 4 generations. Table taken from \cite{Buras:2010cp}.
}
  \label{ratios}
 \end{table}
For example, should observations be made of LFV tau-decays, leading to a ratio   \break
   BR$(\tau\to3 \mu )/$BR$(\tau\to  \mu e e )$ of ${\mathcal O}(1)$, this would disfavour both MSSM cases when compared to the LHT  model (see Table~\ref{ratios}).
Many other such correlations between low-energy cLFV observables could be considered; furthermore, the interplay between low- and high-energy observables (as those that can be studied in colliders) can be also explored. In order to illustrate this, we will focus on the specific case of the SUSY seesaw.

\section{Probing the SUSY seesaw}
The (type I) SUSY seesaw consists of embedding a seesaw mechanism in the framework of SUSY models, which are taken to be flavour conserving, so that the Yukawa couplings are the unique source of flavour violation. Flavour violation in the $\nu$ sector is transmitted to the charged
one via radiative effects involving the $\nu$ Yukawa couplings, $Y^{\nu}$. Even under the
assumption that the SUSY breaking mechanism is flavour conserving,
renormalisation effects (RGE) can induce a sufficiently large  
amount of flavour violation as to account for sizable cLFV 
rates~\cite{Borzumati:1986qx}.
This implies that all cLFV observables will be strongly related. 

At high energies, such as at the LHC, neutralino decays are an excellent laboratory to study cLFV in the slepton sector: among the many possible observables
associated with $\chi_2^0 \to \chi_1^0 \ell_i^\pm \ell_j^\mp$ decays,
one can have sizable flavoured slepton mass splittings, new edges in dilepton invariant mass distributions and explicit flavour violating final states. 
At a future Linear Collider, and in addition to the above observables, one can also study cLFV in 
the $e^\pm e^- \to e^\pm \mu^-$ processes, profiting from the possibility of beam polarization (especially important in reducing charged current LFV background); moreover, the $e^- e^- \to \mu^- \mu^-$ channel might provide a truly golden channel for cLFV.

\subsection{cLFV at LHC and interplay with low-energy cLFV observables}
At the LHC, wino-like neutralino decays into same flavour, opposite
sign dileptons plus missing $E^T$ allow to study cLFV from dilepton
mass distributions. The observables thus considered can also be
correlated with low-energy cLFV, and strategies can be devised to probe the
underlying mechanism of flavour violation in the lepton sector. 

For a flavour conserving framework (for example, the constrained Minimal Supersymmetric SM, cMSSM), and
provided that the SUSY spectrum renders the decays kinematically
viable, one has the following decay chain for a wino-like neutralino:
$\chi_2^0 \to \tilde \ell_{L,R}^i \ell^i \to \chi_1^0 \ell^+_i
\ell^-_i$, mediated by sleptons which are dominated by the flavour
component of the final state lepton. The associated dilepton invariant
mass distributions, $m_{ee}$ and  $m_{\mu \mu}$, exhibit two edges,
corresponding to intermediate left- and right-handed selectrons and
smuons, respectively. The comparison of the $m_{ee}$ and  $m_{\mu
  \mu}$ distributions will further reveal that the edges are
approximately superimposed, corresponding to degenerate sleptons of
the first two generations. This can be inferred from the dotted lines
in Fig.~\ref{edge}, taken from~\cite{Abada:2010kj}.
{\small{
\begin{figure}[h!]
\begin{center}
\raisebox{0mm}{\includegraphics[width=154mm]{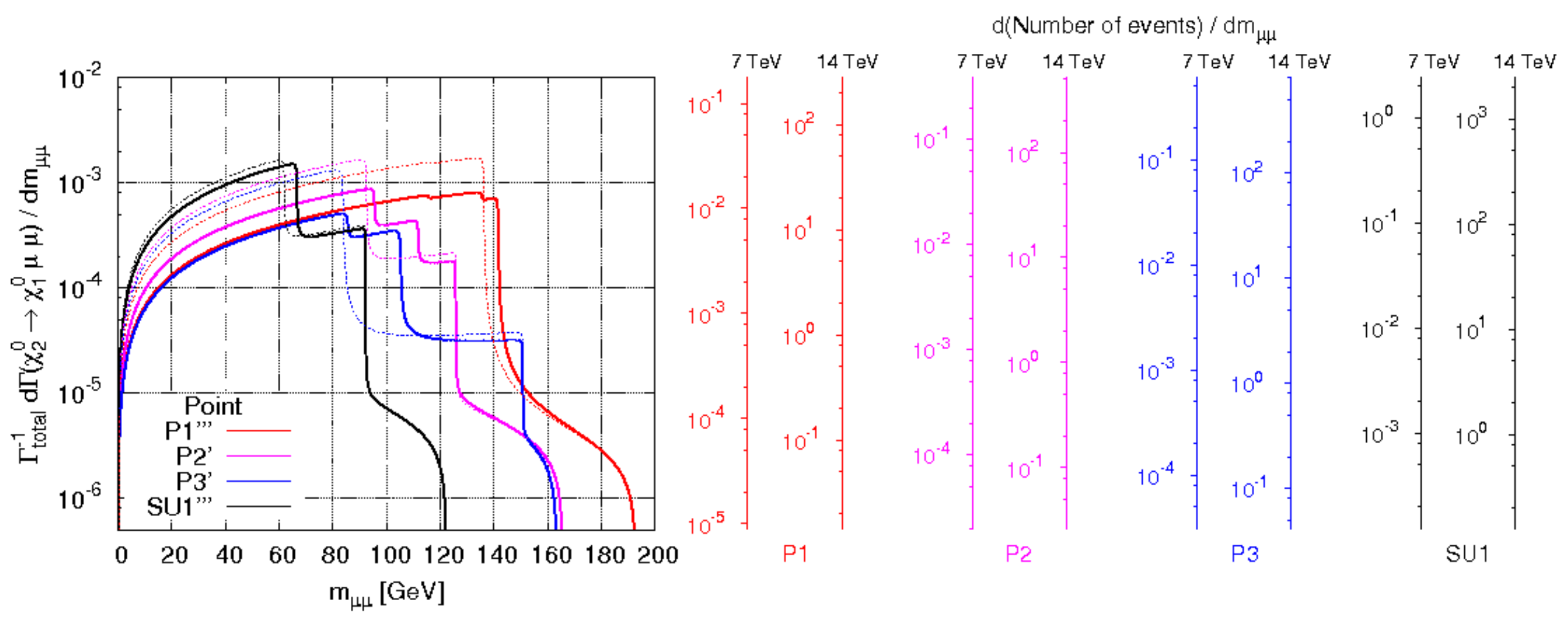}}\vspace*{-1mm}
\caption{\small Evidence of cLFV at high-energies from dimuon invariant mass distributions: BR($\chi_2^0 \to \mu \mu \chi_1^0$) as a function of $m_{\mu \mu}$ for different SUSY seesaw
spectra; a third edge reveals that $\tilde \tau$ also mediated the neutralino decays, in addition to the $\tilde \mu_{L,R}$ characteristic of the flavour-conserving cMSSM case (dotted lines). Secondary y-axes denote the expected number of events for 
$\sqrt s = 7 \ (14)$  TeV,  with 
$\mathcal{L}=1 \ (100)\ \mathrm{fb}^{-1}$.}
\label{edge}
\end{center}
\end{figure}}}
The impact of the SUSY seesaw is visible in the full lines of Fig.~\ref{edge}. 
Firstly, let us mention that should we now compare  $m_{ee}$ and
$m_{\mu \mu}$ distributions we would verify that the edges
corresponding to the left-handed sleptons no longer coincide, which
directly signals that there will be a non-negligible slepton mass
splitting ($\Delta m_{\tilde \ell}/ m_{\tilde \ell} (\tilde e_L,
\tilde \mu_L)$), which can be as large as a few \%.
More striking is the appearance of a third edge in $m_{\mu \mu}$ -
this provides clear and indisputable evidence that a different 
flavour slepton (in this case $\tilde \tau_2$) has mediated the decay,
and signals cLFV in slepton production and decay.

\begin{figure}[h!]
\begin{center}
\begin{tabular}{cc}
\raisebox{0mm}{\includegraphics[width=76mm]{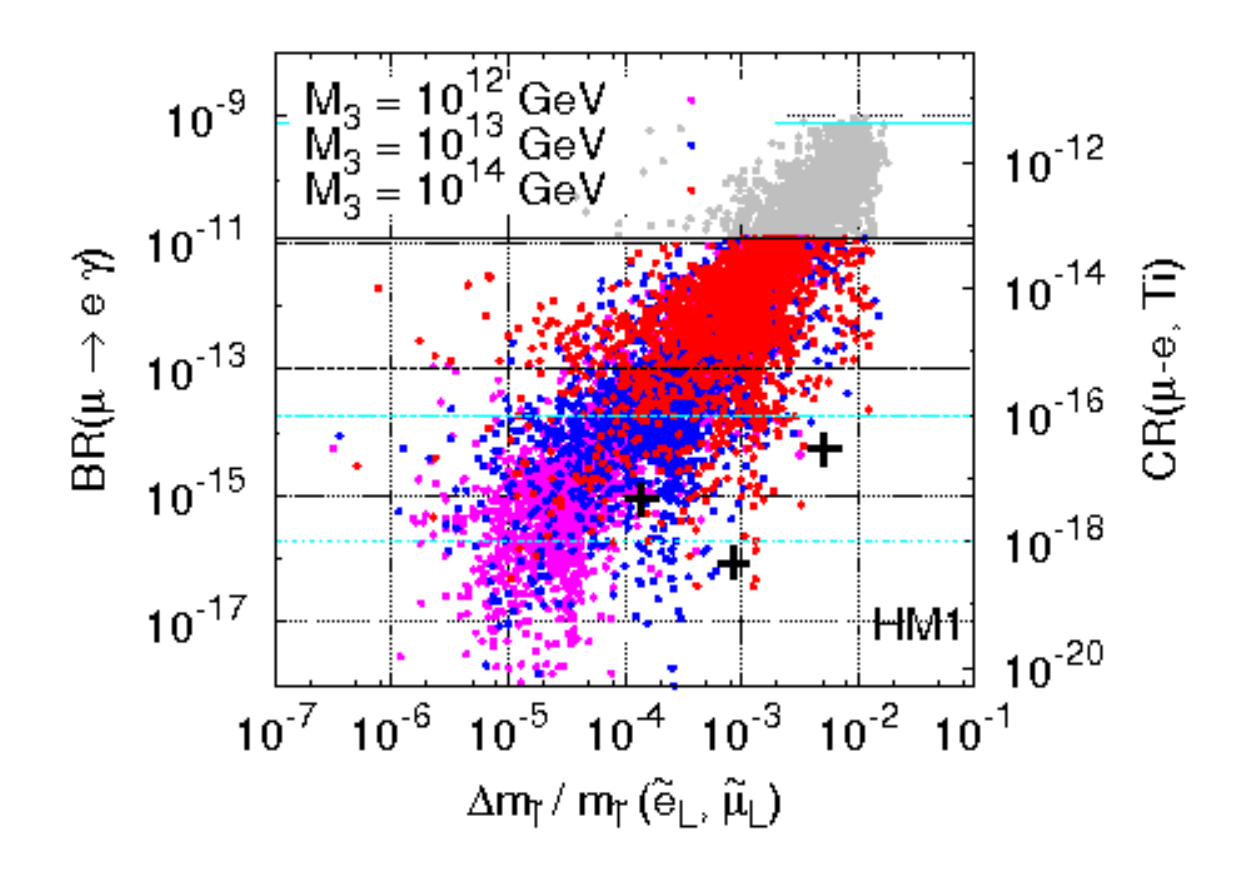}}
&
\raisebox{0mm}{\includegraphics[width=76mm]{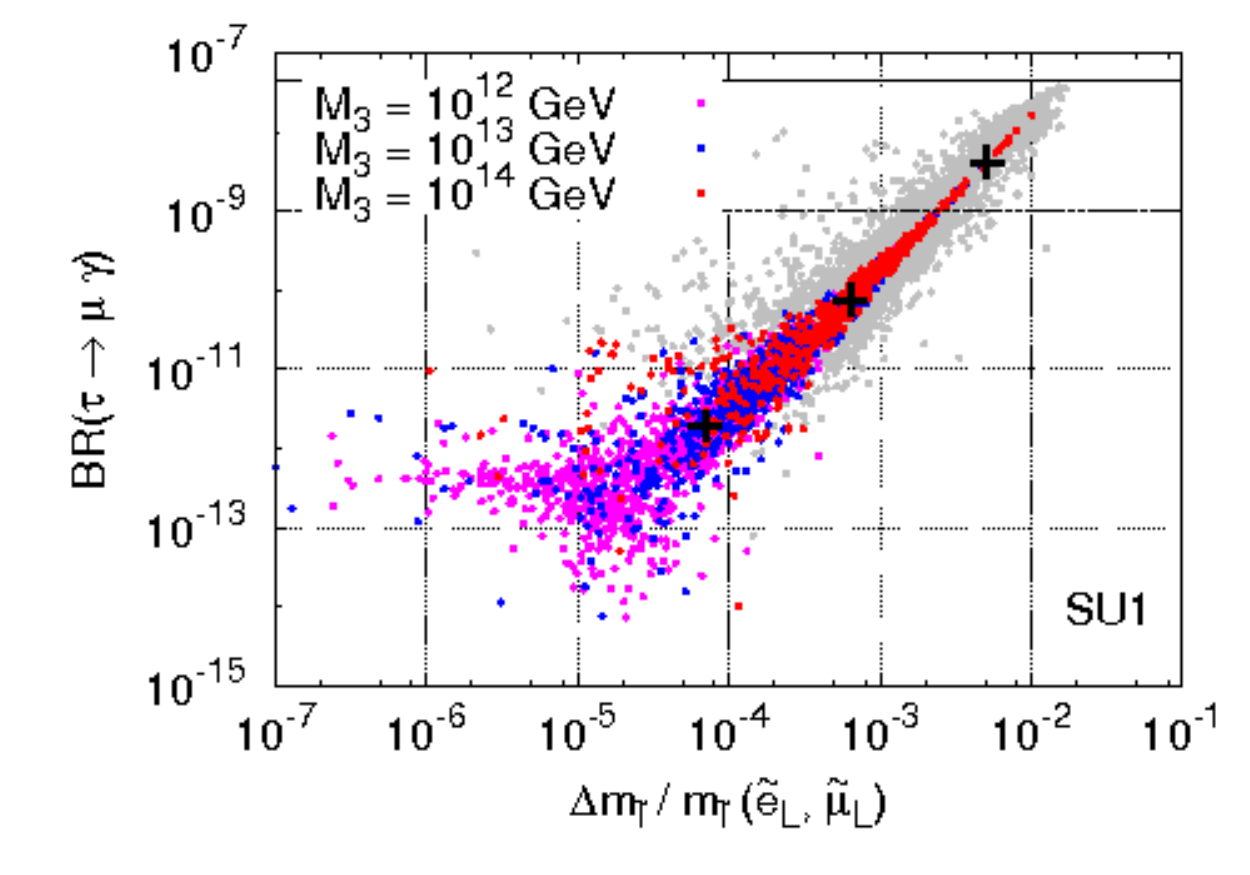}}
\end{tabular}\vspace*{-1mm}
\caption{\small Probing the SUSY seesaw via the synergy of low and high energy cLFV observables: 
on the left BR($\mu \to e \gamma$)  as a function of
 $\tilde e_L - \tilde \mu_L$ mass difference, with corresponding
predictions of CR($\mu-e$, Ti) displayed on the secondary right y-axis; 
on the right BR($\tau \to \mu \gamma$) vs. $\Delta m_{\tilde \ell}/ m_{\tilde \ell} (\tilde e_L,
\tilde \mu_L)$.
Horizontal lines denote the corresponding current bounds/future
sensitivities.}
\label{interplay}
\end{center}
\end{figure}

If a type I SUSY seesaw is indeed at work, then the different
observables are expected to be correlated. Studying the interplay of
different high- and low-energy observables thus allows to test the 
underlying hypothesis of type I seesaw embedded into the cMSSM (or any
other flavour-blind SUSY breaking model). This is illustrated on
Fig.~\ref{interplay}
where we display BR($\mu \to e \gamma$) (with information on the
corresponding ranges for $\mu-e$ conversion in Ti nuclei on the
secondary y-axis), as well as BR($\tau \to \mu \gamma$), all as a
function of the slepton mass splittings. 
The measurement of flavoured mass splittings at the LHC, compatible
with an observation of BR($\mu \to e \gamma$) and/or 
BR($\tau \to \mu \gamma$) would clearly strengthen the seesaw
hypothesis, hinting towards the scale of NP; 
conversely, splittings corresponding to an amount of cLFV 
already excluded by low-energy observables, or the observation of
low-energy cLFV for negligible slepton mass splittings would point towards 
distinct (or additional) sources of  flavour violation. For an updated analysis, see~\cite{Figueiredo:2013tea}.

\subsection{cLFV at Linear Colliders}
\begin{figure}[h!]
\begin{center}
\begin{tabular}{cc}
\raisebox{0mm}{\includegraphics[width=78mm]{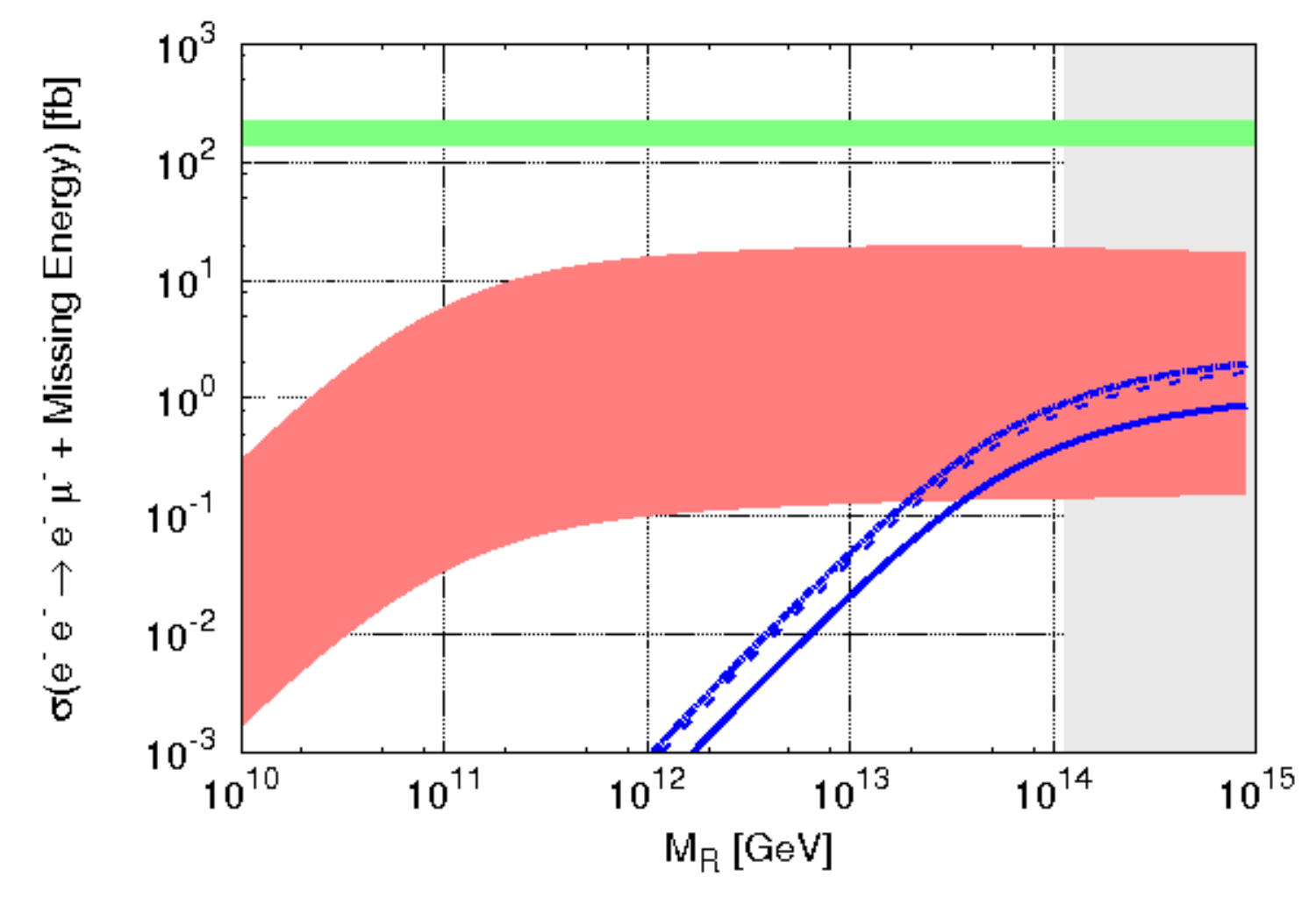}}
&
\raisebox{0mm}{\includegraphics[width=78mm]{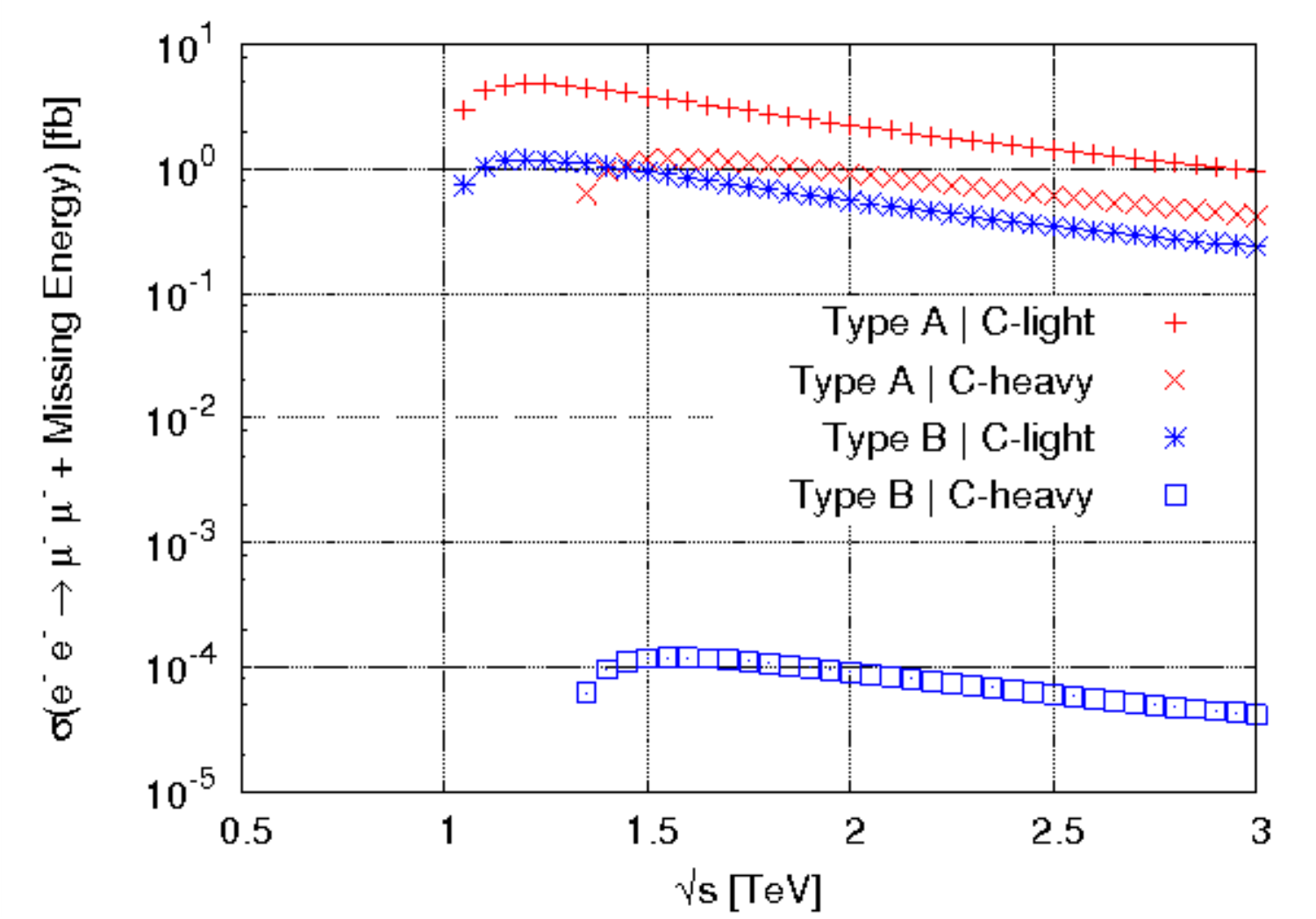}}
\end{tabular}\vspace*{-1mm}
\caption{\small On the left,  cross section for 
$e^- e^- \to e^- \mu^-$ as a function of the seesaw scale, $M_R$; on the right $\sigma(e^- e^- \to \mu^- \mu^-)$ as a function of the c.o.m. energy, for different SUSY seesaw spectra.}
\label{linear}
\end{center}
\end{figure}

A detailed study of cLFV at a future Linear Collider (LC) was conducted in~\cite{Abada:2012re}. When compared to the LHC, a LC offers the possibility of direct slepton production, thus giving access to the slepton sector for  shorter SUSY decay chains.  
In particular, the following processes were studied, 
{\small
\begin{equation}
e^+\,e^-\, \to \,
\left\{
\begin{array}{l}
e^+\,\mu^- + 2\,\chi_1^0 \quad  
\\ 
e^+\,\mu^- + 2\,\chi_1^0 + {(2,4)}\,\nu
\\
e^+\,\mu^- + {(2,4)}\,\nu\end{array}
\right.
\quad \quad
e^-\,e^-\, \to \,
\left\{
\begin{array}{l l}
e^-\,\mu^-
+ 2\,\chi_1^0 \quad 
&{\mathrm{\scriptsize Signal}}\\ 
e^-\,\mu^-
+ 2\,\chi_1^0 +{(2, 4)}\,\nu\quad
&{\mathrm{\scriptsize SUSY\,bkg}}\\
e^-\,\mu^-
+ {(2, 4)}\,\nu\quad 
&\scriptsize{\mathrm{SM}}_{m_\nu}\scriptsize{\mathrm{\,bkg}}
\end{array}
\right.
\end{equation}}
\noindent \hskip -0.15cm corresponding to the SUSY seesaw signal, SUSY and SM$_{m_\nu}$ backgrounds (bkg), respectively.
As can be seen in Fig.~\ref{linear} (left), provided that the seesaw scale is not excessively low, one can have sizable cross sections for 
$e^- e^- \to e^- \mu^-$, above 1 fb (the SM bkg is expected to be disentangled via appropriate cuts, due to the different topology and nature of the missing energy).

In Fig.~\ref{linear} (right) we depict the $e^- e^- \to \mu^- \mu^-$ channel as a function of the centre of mass energy ($\sqrt s$). Provided that $\sqrt s$ is sufficiently high, the signal clearly dominates over the SUSY background (the SM one being negligible). As can be seen, this channel offers a clean probe of the Majorana nature of the exchanged neutral superparticle (neutralino) in the $t$-channel, and can become a truly golden channel to probe the neutrino mass generation mechanism.

\section{Conclusions}
Neutrino oscillation data clearly suggests the existence of NP beyond the SM. However, the actual underlying mechanism of neutrino mass generation and lepton flavour violation remains unclear. Here, we have argued that cLFV observables are a privileged means to address these issues. In particular, the interplay between observables allows to probe different extensions of the SM. For certain frameworks, as is the case of the SUSY seesaw, the possible correlation  between low and high-energy cLFV observables (with the latter being studied at colliders such as the LHC or a future LC) may provide a unique tool to test the source of LFV, further hinting on the typical scale of the mechanism of neutrino mass generation.

\section*{Acknowledgments}
A. A. warmly thanks  the organisers of  the  25th Rencontres de Blois on "Particle Physics and Cosmology" (Blois 2013) and 
 aknowledges support of the ANR project CPV-LFV-LHC NT09-508531 and partial support from the European Union FP7 ITN
INVISIBLES (Marie Curie Actions, PITN- GA-2011- 289442).

\end{document}